\documentclass[11pt,a4paper,headinclude]{scrartcl}

\pdfoutput=1

\usepackage{xcolor}
\definecolor{dark-red}{rgb}{0.4,0.05,0.05}
\usepackage[colorlinks=true,
  bookmarks,
  citecolor=dark-red,
  linkcolor=dark-red]{hyperref}

\usepackage[utf8x]{inputenc}
\usepackage[OT1]{fontenc}
\usepackage[english]{babel}

\usepackage{amsmath}
\usepackage{graphicx}
\usepackage[numbers,compress,square]{natbib}

\usepackage[charter]{mathdesign}
\usepackage{setspace}
\usepackage{isomath} 
\usepackage{titlesec}
\usepackage{tabularx}
\usepackage{caption}
\usepackage{scrpage2}
\usepackage{textcase,soul}
\usepackage{multicol}
\usepackage{hyphenat}

\areaset[0mm]{140mm}{236mm}

\makeatletter
\renewenvironment{table}{\@float{table}\small}{\vskip1ex\end@float}
\renewenvironment{table*}{\@dblfloat{table}\small}{\vskip1ex\end@dblfloat}
\makeatother

\titleformat{\section}
{\bf\centering\large}{}{1em}{}
\titlespacing*{\section}{0pt}{1.2\baselineskip}{0.4\baselineskip}

\titleformat{\subsection}[runin]
{}{}{0em}{\bf}[.\,\,]
\titlespacing*{\subsection}{0pt}{0.7\baselineskip}{0\baselineskip}

\newcommand{\bu}{$\mathrm{\beta}$}
\newcommand{\au}{$\mathrm{\alpha}$}
\newcommand{\bh}{$\boldsymbol{\mathrm{\beta}}$}
\newcommand{\ah}{$\boldsymbol{\mathrm{\alpha}}$}

\newcommand{\ab}{A\bu}
\newcommand{\as}{\au S}

\newcommand{\Lr}{L_\text{r}}
\newcommand{\Fr}{F_\text{r}}

\begin{document}

\thispagestyle{empty}


\vspace*{0cm}
\begin{center}
  {\Huge \noindent \bf Mechanical resistance in \\\vspace{0.4cm}
    unstructured proteins\footnote{Cite as: Biophysical Journal (2013)
      104:2725--2732} }\\
\vspace*{1cm}
{\large\noindent Sigurður Ægir Jónsson\textsuperscript{1}, Simon
  Mitternacht\textsuperscript{2} and Anders Irbäck\textsuperscript{1}
}\\
\end{center}
\vspace*{0.7cm}
\begin{singlespace*}
  {\footnotesize\noindent \textsuperscript{1}Computational Biology \&
    Biological Physics, Department of Astronomy and Theoretical Physics, Lund
    University, Sölvegatan 14A, SE-223 62 Lund, Sweden.
    \textsuperscript{2}University Library, University of Bergen,
    Pb. 7808, N-5020 Bergen, Norway 
  }
\end{singlespace*}
\vspace*{0.5cm}
\begin{small}
\subsection{Abstract}
Single-molecule pulling experiments on unstructured proteins linked to
neurodegenerative diseases have measured rupture forces comparable to
those for stable folded proteins.  To investigate the structural
mechanisms of this unexpected force resistance, we perform pulling
simulations of the amyloid \bu-peptide (\ab) and \au-synuclein (\as),
starting from simulated conformational ensembles for the free
monomers.  For both proteins, the simulations yield a set of rupture
events that agree well with the experimental data.  By analyzing the
conformations right before rupture in each event, we find that the
mechanically resistant structures share a common architecture, with
similarities to the folds adopted by \ab\ and \as\ in amyloid fibrils.
The disease-linked Arctic mutation of \ab\ is found to increase the
occurrence of highly force-resistant structures.  Our study suggests
that the high rupture forces observed in \ab\ and \as\ pulling
experiments are caused by structures that might have a key role in
amyloid formation.\\\\
\noindent \emph{Keywords}: intrinsically disordered proteins,
mechanical unfolding, Monte Carlo simulation, Alzheimer's disease,
Parkinson's disease.
\end{small}

\section{Introduction}
The amyloid-forming proteins linked to neurodegenerative disorders
often belong to the class of intrinsically disordered proteins
\cite{Chiti2006,Knowles:11,Eisenberg2012,Uversky:10}.  Recent
single-molecule atomic force microscopy (AFM) experiments show that,
despite their lack of a well-defined folded structure, such proteins
sometimes exhibit high mechanical resistance when pulled by an applied
force \cite{Sandal2008a, Brucale2009a,Hervas2012a}.  Rupture forces of
similar magnitude are, in fact, only rarely observed for folded
proteins \cite{Valbuena:09,Sikora:11}.  In this paper, we attempt to
convince the reader that subensembles of folded conformations with a
specific architecture can explain this unexpected force resistance.
 
The pathway from a more or less disordered monomer to an ordered
amyloid fibril, with its characteristic cross-\bu\ structure, involves
a host of intermediate assemblies. The exact character of these
species, and their possible roles in pathogenesis, is not completely
understood \cite{Kirkitadze2002,Bemporad2012,Benilova:12}.  It is,
however, clear that at some point the assemblies become rich in
\bu-sheets. Structural motifs, and in particular \bu-sheets, are key
to the mechanical stability of proteins.  In general, the response to
a stretching force is largely dictated by native topology
\cite{Paci:00, Klimov:00} and pulling geometry
\cite{Brockwell:03,CarrionVazquez:03}, and high stability can be
traced to a specific structural unit, a \emph{mechanical clamp}
\cite{Valbuena:09}.  Typically, these units are \bu-sheets, arranged
such that the strands cannot be separated without breaking a number of
hydrogen bonds all at once. The AFM-detected force-resistant states of
disease-related proteins \cite{Sandal2008a,Brucale2009a, Hervas2012a}
therefore appear likely to contain \bu-sheets, and could play a vital
role in the aggregation process.  Knowledge of the structural
specifics of the force-resistant states could thus provide valuable
and general insight into the aggregation mechanisms in amyloid
diseases.

In this study, we use atomic-level Monte Carlo (MC) simulations to
search for and characterize possible mechanical clamps in Alzheimer's
amyloid \bu-peptide (A\bu) and the Parkinson's disease-linked
\au-synuclein protein (\au S). Both proteins are unstructured and
amyloid-forming, and displayed mechanical resistance in AFM
experiments~\cite{Sandal2008a, Brucale2009a,Hervas2012a}.
Computational modeling offers unique opportunities to interpret this
force response, but requires the use of large sets of trajectories,
due to the structural diversity of the proteins. Our calculations are
carried out starting from large and diverse sets of initial
conformations, randomly drawn from previously simulated
ensembles~\cite{Mitternacht2010, Jonsson2012a}.  In the simulations,
high-force rupture events occur for both proteins. We find that these
events are caused by one specific type of mechanical clamp, which
shows similarities to the folds adopted by
\ab~\cite{Petkova:02,Luehrs:05} and \as~\cite{Vilar:08} in fibrils.
Such structures could play an important role in the addition of
monomers to growing \ab\ and \as\ fibrils, an association that
requires large parts of the monomers to adopt a specific fold.  In
addition to the wild-type (WT) sequences, we study the disease-related
Arctic E22G mutant \cite{Nilsberth:01} of \ab, and find it to increase
the occurrence of strong mechanical clamps.

\section{Methods}

\subsection{Model} 
Our simulations are based on an all-atom implicit solvent model with
torsional degrees of freedom \cite{Irback2009}.  The energy function
consists of four main terms
\begin{equation*}
  E = E_{\text{ev}} + E_{\text{loc}} + E_{\text{hb}} + E_{\text{sc}}\,.
\end{equation*}
The $E_{\text{ev}}$ term represents excluded volume effects, whereas
$E_{\text{loc}}$ handles local interactions among neighboring
atoms. Hydrogen bonds between backbone NH and CO groups, and between
charged side-chain groups and the backbone are handled by
$E_{\text{hb}}$. The last term, $E_{\text{sc}}$, represents
interactions between side chains, both hydrophobic attraction and
attraction/repulsion between charged side chains. The mathematical
form and parameters of the model are given elsewhere
\cite{Irback2009}.  The computational convenience of the model allows
us to generate large sets of trajectories, which is important because
of the conformational polymorphism of the proteins studied.

One of the big challenges in creating protein models is to obtain a
realistic temperature dependence.  The temperature scale of our model
was determined to give correct folding temperatures for a diverse set
of small peptides~\cite{Irback2009}. Our simulations have, however,
shown that the temperature scale needs recalibration for the larger
\ab\ and \as\ proteins~\cite{Mitternacht2010,Jonsson2012a}.  Below we
will both give the nominal temperatures used in the simulations and
indicate what physical temperatures these correspond to, based on
comparisons with experimental data.

\subsection{Pulling}

We model the action of the pulling force by adding a harmonic
potential in  the end-to-end distance of the protein, $L(x)$ (between
the first and last backbone atoms), where $x$ denotes a protein
conformation. The equilibrium length is initially set to the
end-to-end distance of the starting conformation, $L_0 = L(x_0)$, and
then increases with a constant velocity $v$. The full energy function
is given by
\begin{equation}
  E_{\text{tot}}(x,t) = E(x) + \frac k 2  \left[L_0+vt-L(x)\right]^2,
\end{equation}
where $E(x)$ is the energy in the absence of external force, $t$ is MC
time and $k$ a spring constant. In our calculations, we set $k =
37\,\mathrm{pN/nm}$ and $v=0.05$\,fm/MC step.  The AFM study of
\ab\ and \as\ by Herv\'as et al. \cite{Hervas2012a} was performed
using an estimated cantilever stiffness of 30--70\,pN/nm and a pulling
speed of 0.4\,nm/ms. Our $v$ cannot be directly translated to an
experimental pulling speed, but was empirically chosen based on
previous simulations of fibronectin domains
\cite{mitternacht:09}. Here, we found simulation results obtained
using $v=0.1$\,fm/MC step to be compatible with AFM measurements at a
pulling velocity of 0.6\,nm/ms. Therefore, in our present
calculations, we expect rupture forces comparable to those measured by
Herv\'as et al. \cite{Hervas2012a}, especially since the velocity
dependence is only logarithmic \cite{Evans:97}.  Generating \ab\ and
\as\ unfolding trajectories at experimental pulling velocities by
conventional molecular dynamics methods is a challenge, as the
required time span of each trajectory is $\sim$\,0.1\,s.

\subsection{MC details}
We simulate the above model using MC dynamics. MC is a crude
approximation to the real dynamics over short time scales, but we
expect MC-based unfolding simulations to capture the major free-energy
minima and barriers encountered by a protein when stretched, provided
that only small trial moves are used \cite{Tiana2007,Vitalis:09}.  Our
calculations are based on two ``small-step'' MC updates: rotations of
single side-chain angles and coordinated Biased Gaussian Steps (BGS)
for backbone angles \cite{Favrin:01}. BGS updates up to eight
consecutive backbone torsion angles in a manner that keeps the ends of
the updated segment approximately fixed. The equilibrium length of the
harmonic spring pulling the protein is increased by a small constant
amount in each MC step.

\subsection[A-beta ensembles]{A\bh\ ensembles}
We study the highly amyloidogenic 42-residue form of \ab.  In previous
work, we investigated monomer properties and dimer formation for four
variants of this peptide: WT, E22G, F20E and E22G/I31E
\cite{Mitternacht2010,Mitternacht:11}. The model temperature
(nominally $37\,^{\circ}\mathrm{C}$) was chosen for best match with
experimental chemical shifts \cite{Hou:04} and $J$-couplings
\cite{Sgourakis:07} measured at 0--5$\,^{\circ}\mathrm{C}$. The
present study is conducted at the same temperature, and focuses on the
WT and E22G variants. To ensure proper sampling, new extended
simulations of the free monomers were performed, which indeed agree
very well with previous data. Starting from conformations randomly
drawn from these equilibrium ensembles, a set of 512 independent
pulling trajectories was generated for each of WT \ab\ and E22G \ab.
  
\subsection[alpha-S ensemble]{\ah S ensemble}
In our previous study of the free 140-residue \as\ monomer, we
observed two distinct phases: a disordered high-energy phase, D, and a
\bu-structure-rich low-energy phase, B \cite{Jonsson2012a}.  The
transition frequency between these phases was insufficient for a
determination of their relative fraction as a function of temperature.
Our analysis therefore focused on single-phase properties, evaluated
using a fixed model temperature (nominally $56\,^{\circ}\mathrm{C}$)
at which both phases coexisted. The generated D phase ensemble was
found to be in good agreement with NMR data at
$15\,^{\circ}\mathrm{C}$ \cite{Eliezer:01}.  A consistent description
of NMR data at $-15\,^{\circ}\mathrm{C}$ and $-10\,^{\circ}\mathrm{C}$
\cite{Kim2007,Wu2008} in terms of the B and D phases was, by contrast,
found to require a significant B phase fraction, around 50--70\,\%.
Our pulling simulations are carried out using the same model
temperature ($56\,^{\circ}\mathrm{C}$), with initial conformations
randomly drawn from the two single-phase equilibrium ensembles (total
number 1024). As in the previous study, the two phases are analyzed
separately.

\subsection{Analysis} 
\begin{figure}
  \centering
  \includegraphics[width=8cm]{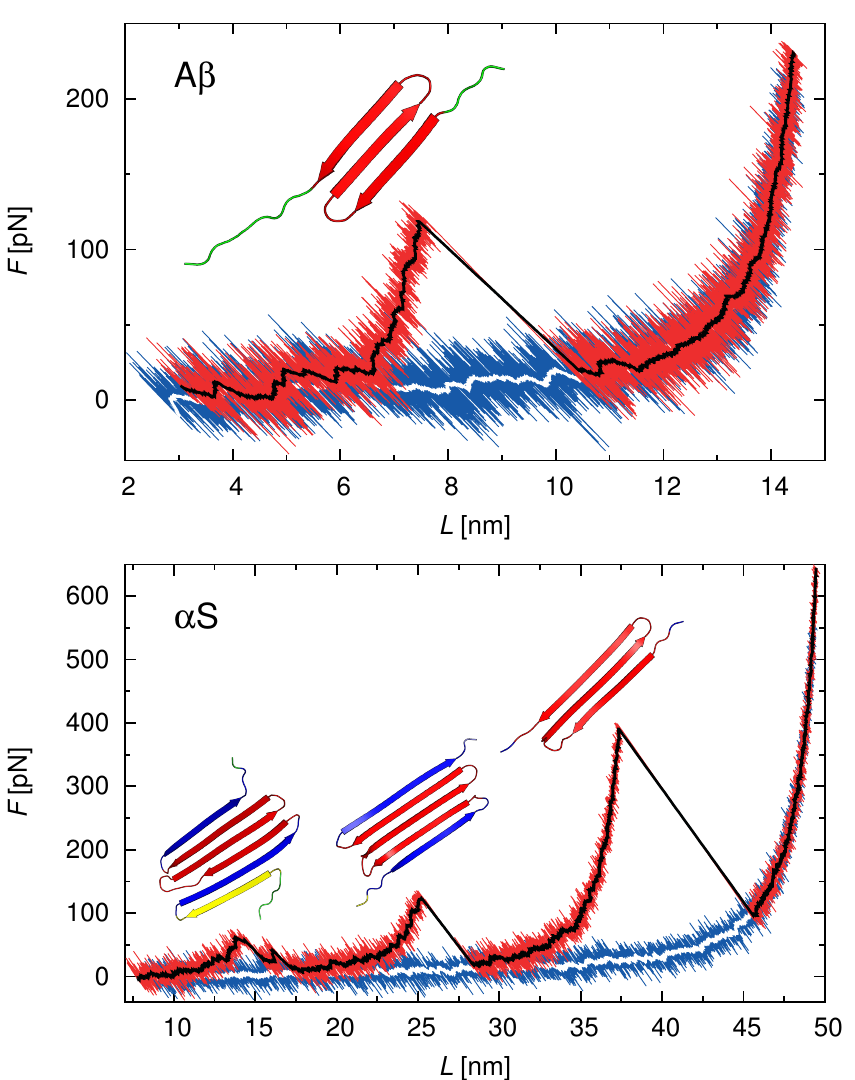}
  \caption{Examples of simulated unfolding trajectories.  (\emph{Upper
      panel}) Force, $F$, versus end-to-end distance, $L$, in two runs
    for WT \ab. In one trajectory, no force-resistant state is
    encountered as the molecule is stretched.  In the other, a rupture
    event occurs at $L_\text{r}=7.5$\,nm and $F_\text{r}=119$\,pN. The
    three-dimensional structure is a snapshot just before the rupture
    event.  The black and white curves illustrate the noise-filtering
    used when analyzing the data (see Methods).  (\emph{Lower panel})
    Similar plot showing two \as\ trajectories with zero and three
    rupture events, respectively. The snapshots illustrate the core
    structures involved in the rupture events.  The first snapshot
    shows residues 1--101, whereas the final three-stranded \bu-sheet
    is formed by the 33--85 segment.
    \label{fig:1}}
\end{figure}

In the pulling simulations, we monitor the external force, $F$, and
the end-to-end distance, $L$. Some examples of force-distance
trajectories can be found in Fig.~\ref{fig:1}. In our analysis of the
generated trajectories, the data are noise filtered using a sliding
window of size $T_\text{w}=6\times10^6$ MC steps. We identify rupture
events as sudden drops in force with MC time. To discriminate true
rupture events from noise, we require the force to drop by $>$\,20\,pN
within a time interval of $T_\text{w}$, from a peak value $>$\,20\,pN
(the force can drop to negative values). For each rupture event, we
register the maximum force, $\Fr$, and the end-to-end distance, $\Lr$,
at which the peak occurs.

Secondary structure is classified using the program \textsc{stride}
\cite{stride}.  The \bu-structure content is calculated as the
fraction of residues that are in the class \emph{extended}. Solvent
accessibility calculations are performed using the program
\textsc{naccess} \cite{naccess}.

To locate \bu-hairpin turns, a contact-based measure of turn
propensity, $\tau_i$, is calculated. Two residues are said to be in
contact if two or more heavy-atom pairs are within 4.5\,\AA\ of each
other. The measure $\tau_i$ is defined as the sum of all contact
probabilities $p_{jk}$ in a strip perpendicular to the main diagonal
of the contact map, $2(i-1)<j+k<2(i+1)$.

Figures of three-dimensional structures are drawn using \textsc{pymol}
\cite{pymol}.


\section{Results}

\subsection[WT A-beta]{WT A\bh} 
The \ab\ molecule is small and natively unfolded, and may seem
unlikely to exhibit any mechanical resistance. However, in their AFM
study, Herv\'as et al. \cite{Hervas2012a} found that many unfolding
traces contained at least one force peak $>$\,20\,pN. The maximum
recorded rupture force exceeded 300\,pN.

\begin{figure}
  \centering
  \includegraphics{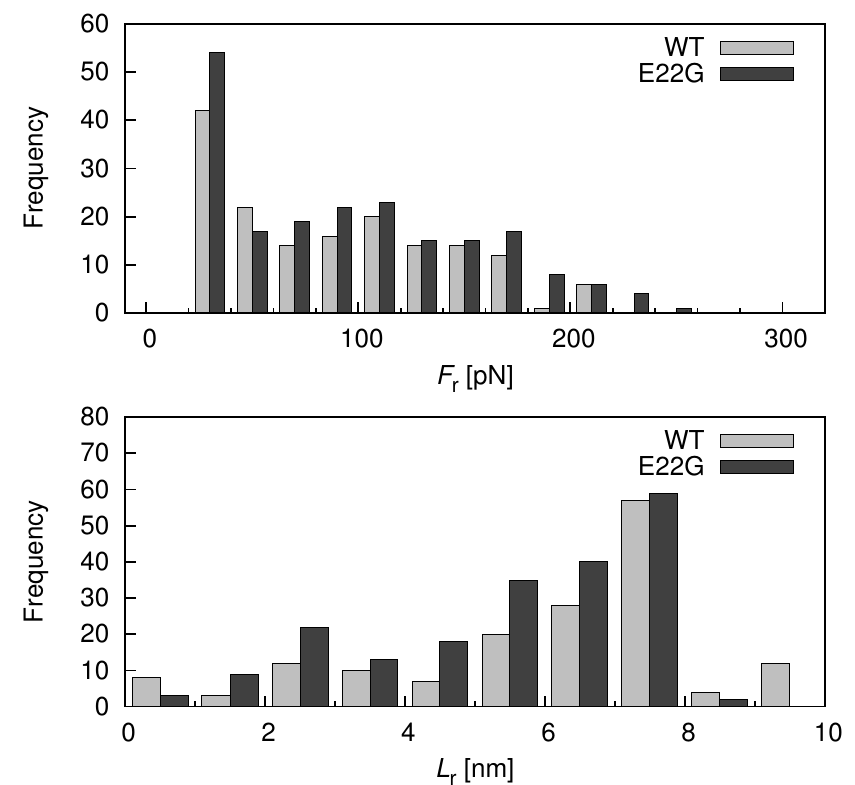}
  \caption{ Histograms of rupture force (\emph{upper panel}), $\Fr$,
    and rupture distance (\emph{lower panel}), $\Lr$, for WT and E22G
    \ab, based on 161 and 201 rupture events, respectively.
    \label{fig:2}}
\end{figure}

We examine the structural mechanisms causing this force resistance
using MC simulations.  As described in Methods, starting from a
simulated ensemble of conformations, we study the response of the
protein when pulled at constant velocity.  In some of the simulated
trajectories, when stretched, the \ab\ molecule gets trapped in a
force-resistant state, the breaking of which gives rise to a force
maximum (Fig.~\ref{fig:1}).  Rupture events, signaled by force peaks
$>$\,20\,pN, occur in 148 of our 512 unfolding trajectories (see
Methods for a description of our peak detection protocol). Most of
these 148 runs contain one rupture event, but there are also
trajectories with two such events (Table~\ref{tab:1}).
\begin{table}
  \centering
  \begin{tabular}{llll} 
    Protein & $N_\text{traj}$ & $N_\text{mech}$ & $N_\text{ev}$\\ \hline
    \ab\ WT & 512 & 148 (135, 13, 0, 0, 0) & 161\\
    \ab\ E22G & 512 & 172 (144, 27, 1, 0, 0) & 201 \\
    \as, D phase& 449 & 2 (1, 1, 0, 0, 0) & 3 \\
    \as, B phase& 575 & 563 (37, 205, 274, 43, 4) & 1461 \\
    \hline
  \end{tabular}
  \caption{Rupture event statistics. The total
    number of trajectories ($N_\text{traj}$), the number of
    trajectories with at least one rupture event ($N_\text{mech}$),
    and the total number of observed rupture events ($N_\text{ev}$)
    for the four systems studied.  The numbers of trajectories with 1,
    2, 3, 4 and 5 rupture events are indicated within parentheses in
    the $N_\text{mech}$ column.}
  \label{tab:1}
\end{table}
For each rupture event, we record the maximum force, $\Fr$, and the
end-to-end distance, $\Lr$, at which the peak occurs.  The
distributions of $\Fr$ and $\Lr$ are broad (Fig.~\ref{fig:2}), showing
that the observed rupture events are caused by a range of structures
with varying mechanical strength.
Our simulations differ in several details from experiments, where, for
instance, the molecule of interest is embedded in a multimodular
construct.  Despite these uncertainties, our results are in good
overall agreement with those reported by Herv\'as et
al.~\cite{Hervas2012a}. The conformational polymorphism that we
observe resembles that found in the experiments, and our measured
rupture forces are comparable to the experimental ones.
   
\begin{figure}
  \centering
  \includegraphics{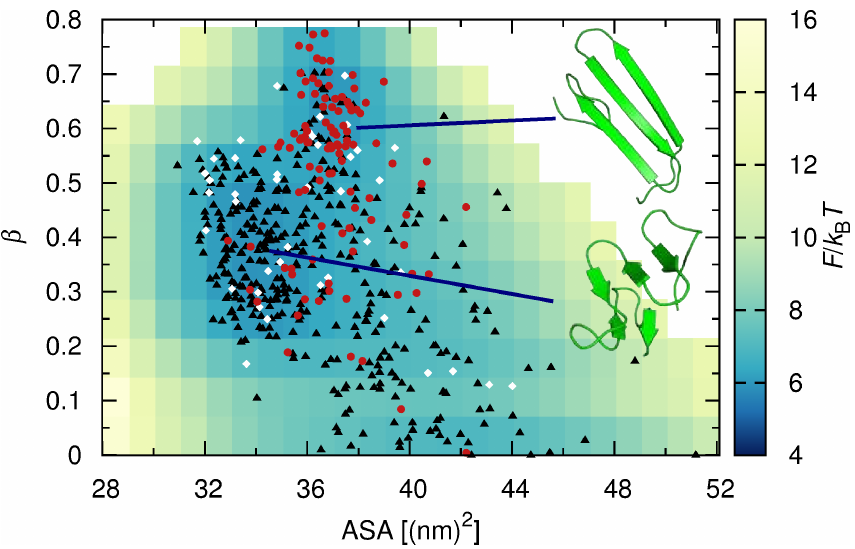}
  \caption{Equilibrium free energy, $F(\text{ASA},\beta)$, calculated
    as a function of accessible surface area, ASA, and \bu-structure
    content, $\beta$, for WT \ab. Plot symbols represent starting
    structures for the pulling simulations, using averages of ASA and
    $\beta$ over a short time interval ($6\times 10^6$ MC steps).  The
    plot symbol indicates whether the maximum rupture force along the
    corresponding pulling trajectory is $<$\,20\,pN (triangle),
    20--150\,pN (diamond), or $>$\,150\,pN (circle).  The free energy
    has two shallow minima. The three-dimensional structures represent
    these minima.\label{fig:3}}
\end{figure}

Having seen this agreement, we next examine the structures causing the
simulated rupture events. What conformational features give rise to
rupture forces as high as 210\,pN (Fig.~\ref{fig:2})? The starting
conformations for our pulling simulations are drawn from a diverse
equilibrium ensemble, generated as described in Methods.
Fig.~\ref{fig:3} shows the free energy of this ensemble, calculated as
a function of accessible surface area and \bu-structure content.
Two major free-energy minima can be identified, both shallow and
broad. One minimum corresponds to single extended \bu-sheets and the
other to more compact, often double-layered structures, with lower
accessible surface area ($<$\,36\,nm$^2$) and lower \bu-structure
content ($<$\,0.5).  The plot symbols in Fig.~\ref{fig:3} represent
the conformations from which our pulling simulations are started, and
indicate the magnitude of the maximum rupture force along the
respective trajectories.  A majority of the trajectories containing
rupture events start out from the free-energy minimum corresponding to
single extended \bu-sheets. Conformations from the other minimum have
shorter \bu-strands and are typically unable to withstand high forces.

\begin{figure}
  \centering
   \includegraphics{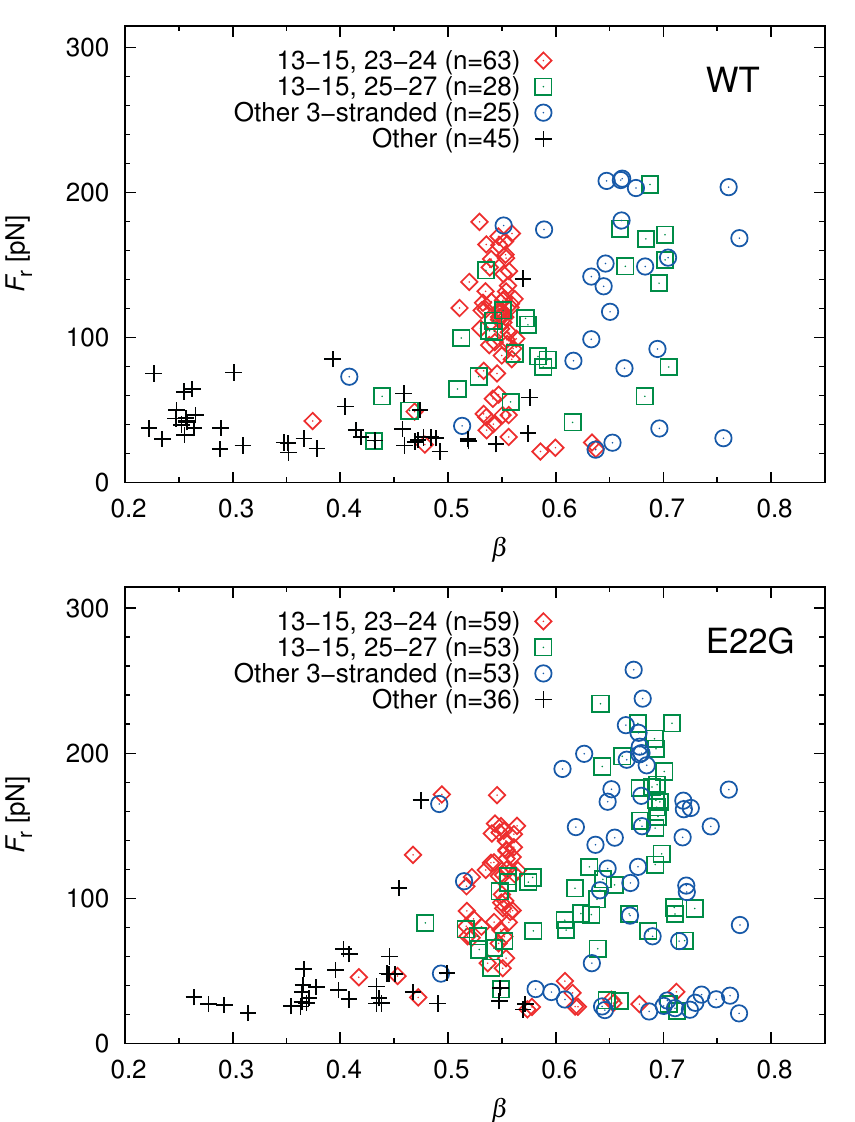}
   \caption{Scatter plots of rupture force, $\Fr$, and \bu-structure
     content, $\beta$, for WT (\emph{upper panel}) and E22G (\emph{lower
       panel}) \ab.  $\beta$ is an average over a short time interval
     ($6\times 10^6$ MC steps) just before the rupture event. The plot
     symbols indicate four conformational classes.  The first two
     correspond to three-stranded \bu-sheets with specific turn
     locations. The third class represents extended three-stranded
     \bu-sheets with other turn locations. The ``other'' category
     consists of structures with at most two extended \bu-strands,
     some of which are double-layered.
    \label{fig:4}}
\end{figure}

To pinpoint the structural elements providing mechanical resistance,
we examine the conformations sampled just before the rupture events.
An analysis of secondary-structure content before rupture shows that
high rupture forces occur only if the \bu-structure content is high
(Fig.~\ref{fig:4}).
For a more detailed picture, we divide the rupture events into four
classes based on \bu-sheet geometry, as indicated by the plot symbols
in Fig.~\ref{fig:4}. The first two classes correspond to
three-stranded \bu-sheets with specific turn locations (turns are
identified using the measure $\tau_i$ defined in Methods). The turns
are located in the 13--15 and 23--24 intervals for the first class,
and in the 13--15 and 25--27 intervals for the second class.  The
third class represents all three-stranded \bu-sheets with other turn
locations. All these three classes contain many high-$\Fr$ events, and
the average $\Fr$ is $>$\,100\,pN in all three cases.  The fourth and
last class collects all the remaining rupture events. The
corresponding structures are often double layered, and sometimes
contain smaller \bu-sheets. For this class, the average rupture force
is $\sim$\,40\,pN.  This analysis confirms that most of the main
mechanical clamps are three-stranded \bu-sheets. Typically, they have
a simple meander topology. \bu-sheets with an odd number of strands in
a meander pattern are indeed obvious candidates for mechanical
clamps. Unlike a \bu-hairpin, they cannot be unzipped one hydrogen
bond at a time.

\begin{figure}
  \centering
   \includegraphics[width=8cm]{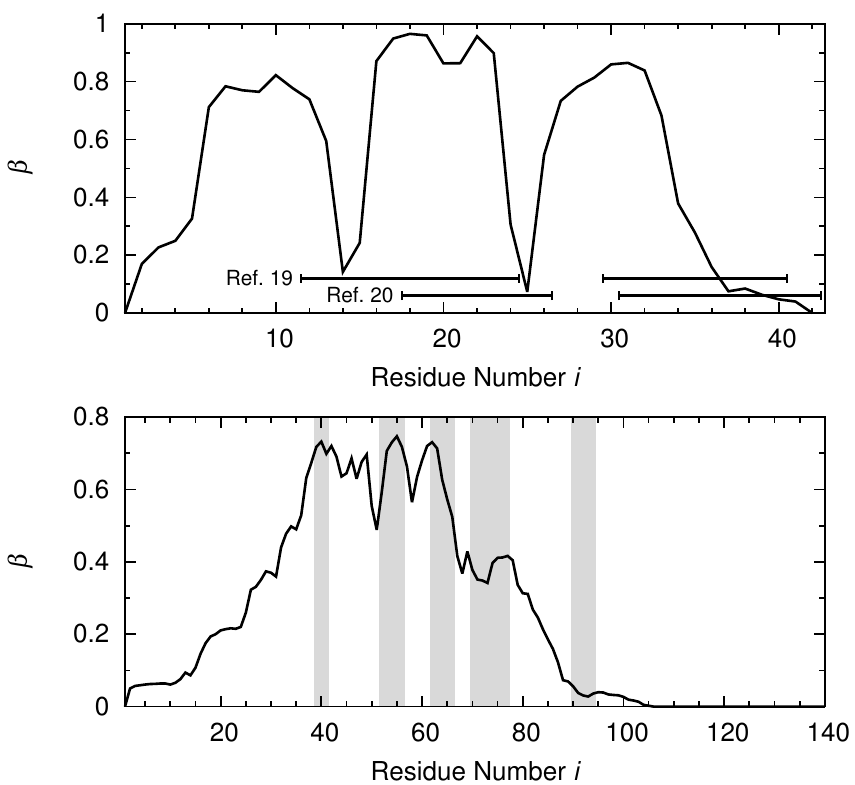}
  \caption{Average \bu-structure profiles for WT \ab\ (\emph{upper
      panel}) and \as\ (\emph{lower panel}) rupture events involving
    three-stranded \bu-sheets. For each event, residue-specific
    \bu-structure propensities are computed over a short time interval
    ($6\times 10^6$ MC steps) just before rupture. A second average
    over events gives the final profile.  Horizontal bars in the upper
    panel indicate strand regions in \ab\ fibrils as obtained by
    Petkova et al.~\cite{Petkova:02} and L\"uhrs et
    al.~\cite{Luehrs:05}.  In the lower panel, \as\ fibril strand
    regions~\cite{Vilar:08} are shaded gray.
  \label{fig:5}}
\end{figure}

The three-stranded \bu-sheets causing rupture events share similar
sequence locations, as one might expect for a small protein such as
\ab.  To quantify this, we show in Fig.~\ref{fig:5} a
$\beta$-structure profile, obtained by averaging over structures just
before rupture in all events involving three-stranded \bu-sheets. The
profile shows that there indeed are statistically preferred turn and
strand regions.  In addition, it can be seen that the second and third
strand regions overlap with the two strand regions in
\ab\ fibrils~\cite{Petkova:02,Luehrs:05}.

For a \bu-sheet to display mechanical resistance, its strands must be
sufficiently long. In the case of \ab, the required strand length can
be achieved only if a large part of the whole molecule participates in
the \bu-sheet. In fact, among the observed force peaks $>$\,100\,pN
for the two \ab\ variants, almost all are caused by structures with a
\bu-structure content $>$\,0.5 (Fig.~\ref{fig:4}). By forming such
structures, it becomes possible for \ab, despite its small size, to
withstand these high forces.

It is worth noting that these \bu-sheets still do not have maximal
size. In WT \ab\ simulations at a lower temperature, we observed even
larger three-stranded \bu-sheets (\bu-structure contents of around
0.8). In pulling simulations started from these conformations, we
recorded rupture forces of up to 300\,pN.  However, these
conformations are not significantly sampled at the temperature used in
the present study.

\subsection[Arctic A-beta]{Arctic A\bh}
Having identified candidate structures for mechanical resistance, it
is still not clear what relevance these have for aggregation and
disease. A natural next step is to study how the force response is
altered by disease-related mutations, such as the (Arctic) E22G
mutation. The experiments by Herv\'as et al. \cite{Hervas2012a} found
this mutation to lead to an increased occurrence of force-resistant
states. The highest force peak was $>$\,400\,pN, indicating the
breaking of a ``hyper-mechanostable'' conformation.
 
We study E22G \ab\ using the same setup as for WT \ab, and see many
similarities between the two variants, as is expected from a single
point mutation, but also some important differences.  Compared to WT,
we observe 25\,\% more rupture events for E22G and 15\,\% more
mechanically resistant initial conformations (Table~\ref{tab:1}). In
particular, there is an increase by 60\,\% in the number of events
with $\Fr>150$\,pN for E22G (40 vs 25). Our highest $\Fr$ is
$\sim$\,260\,pN for this variant, compared to $\sim$\,210\,pN for WT
(Fig.~\ref{fig:2}).

The increased number of high-$\Fr$ events for the E22G mutant can, at
least partly, be traced to a conformational difference noted in our
previous work \cite{Mitternacht2010,Mitternacht:11}. Here, we compared
the propensities of the four variants WT, E22G, F20E and E22G/I31E to
form turns centered in the 25--30 interval, which is the loop region
of the \bu-loop-\bu\ motif in \ab\ fibrils
\cite{Petkova:02,Luehrs:05,Ahmed:10}.  The probability for such turns
was found to correlate with the rate of fibril formation, and was, in
particular, higher for E22G than for WT
\cite{Mitternacht2010,Mitternacht:11}. If the \ab\ molecule is to form
a large and potentially force-resistant three-stranded $\beta$-sheet,
the 25--30 region is a favorable location for the second turn.

\begin{figure}
  \centering
  \includegraphics{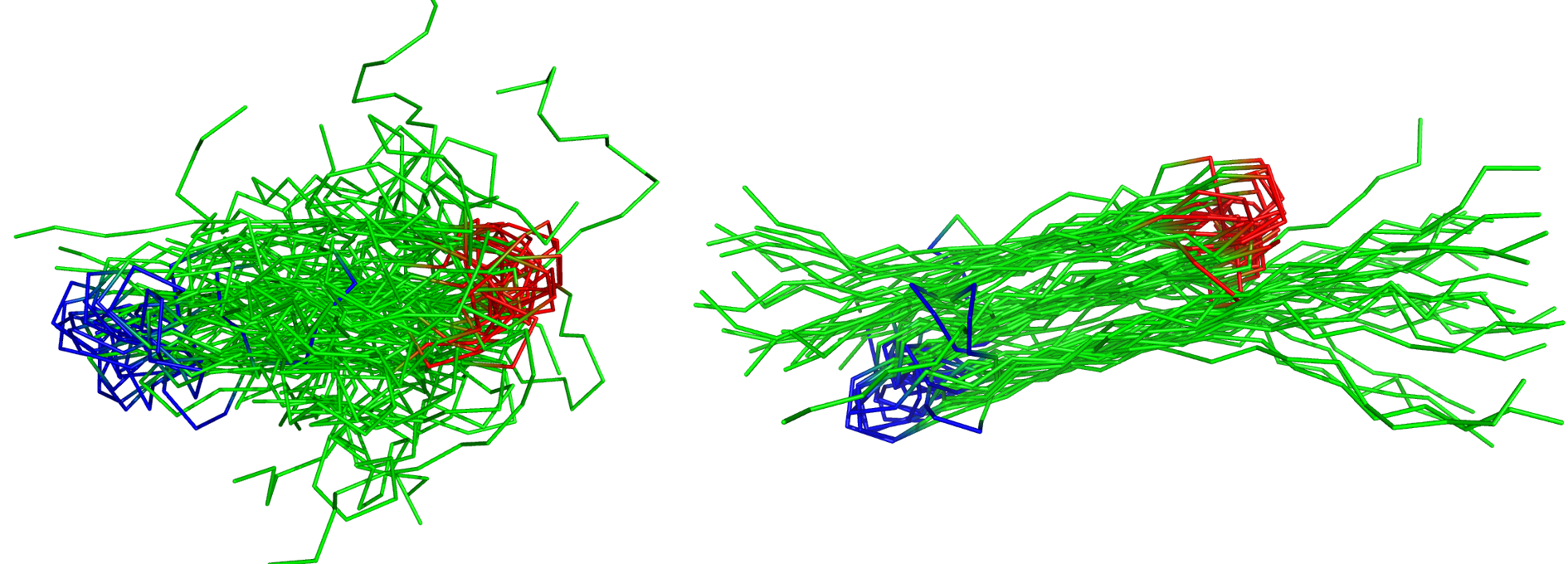}
  \caption{ Aligned WT \ab\ structures before pulling (\emph{left})
    and at rupture (\emph{right}) for the 28 observed rupture events
    in the class with turns at 13--15 and 25--27 (squares in
    Fig.~\ref{fig:4}). For the initial structures residues 10--35 were
    aligned, for the other case residues 5--35. Residues 13--15 are
    in red and residues 25--27 in blue in both alignments.
    \label{fig:6}}
\end{figure}

The high-$\Fr$ events for E22G can to a large extent be linked to
three-stranded \bu-sheets with \bu-structure contents of $\sim$\,0.7
(Fig.~\ref{fig:4}). Many of the conformations are in the class with
the turns located in the 13--15 and 25--27 regions.  In line with the
above-mentioned analysis \cite{Mitternacht2010,Mitternacht:11}, we
find that the number of rupture events in this conformational class
increases from 28 for WT to 53 for E22G. Fig.~\ref{fig:6} shows
aligned initial conformations and conformations right before rupture
for this class, illustrating how the simulations start from a diverse
set of conformations which ``condense'' to one single structure when
pulled.

\subsection[alpha-S]{\ah S}
AFM experiments by different groups have shown that the Parkinson's
disease-related \as\ protein can display mechanical resistance
\cite{Sandal2008a,Brucale2009a,Hervas2012a}. The set of around 100
pulling trajectories generated and analyzed by Herv\'as et
al.~\cite{Hervas2012a} included two in the hyper-mechanostable
category ($\Fr>400$\,pN).

Our \as\ study is based on a set of 1024 constant-velocity pulling
simulations, two of which are shown in Fig.~\ref{fig:1}. As in the
\ab\ study, initial conformations are randomly drawn from a simulated
ensemble for the free monomer \cite{Jonsson2012a}.  The \as\ ensemble
can be split into two phases: a disordered phase, called D, and a
\bu-structure-rich phase, B (see Methods for a detailed
description). Because the phases are very different, we analyze the
two subsets of trajectories separately.

We find that the D phase displays virtually no force resistance.
Force peaks $>$\,20\,pN occur in only 2 of our 449 trajectories
started from this phase. In both these runs, the initial energy is
close to the cutoff used to distinguish the phases. Among our 575
trajectories started from the B phase, there are, in contrast, only 12
without any force peak $>$\,20\,pN.  The typical B phase conformation
exhibits one large \bu-sheet with several strands, often arranged in a
simple meander pattern \cite{Jonsson2012a}.  When stretched, it breaks
up through a sequence of usually two or three rupture events
(Table~\ref{tab:1}), at each of which a few strands are lost
(Fig.~\ref{fig:1}).  Force peaks occurring later in time tend to be
higher than earlier ones, at least in part because the strands can be
more parallel to the applied force as they get fewer in number.  In
most runs ($>$\,90\,\%), the last force peak is associated with the
breaking of a three-stranded \bu-sheet.

\begin{figure}
  \centering
  \includegraphics{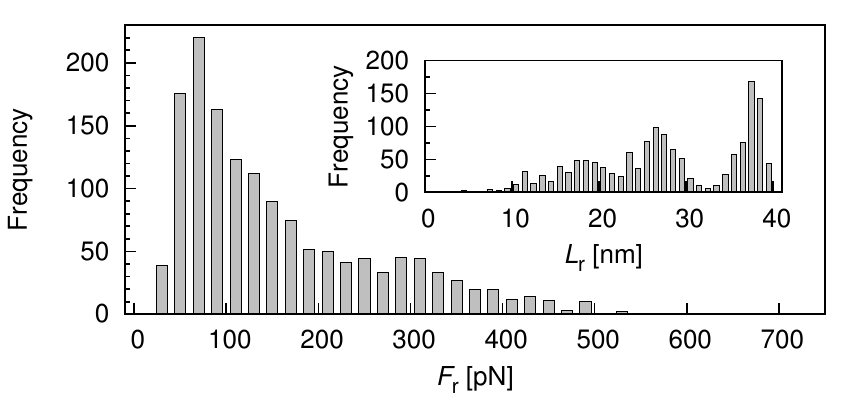}
  \caption{ Histograms of rupture force, $\Fr$, and rupture distance
    (\emph{inset}), $\Lr$, for \as, based on a total of
    1464 rupture events, the vast majority
    (1461) of which are from trajectories started from
    the B phase.\label{fig:7}}
\end{figure}

In accord with the observation of conformational polymorphism in AFM
experiments \cite{Hervas2012a}, our in total 1464 \as\ rupture events
show a wide variation in both $\Fr$ and $\Lr$ (Fig.~\ref{fig:7}).  The
$\Fr$ distribution starts to fall off at $\sim$\,300\,pN. The maximum
$\Fr$ is $\sim$\,740\,pN, but there are only seven events with
$\Fr>500$\,pN. The $\Lr$ distribution exhibits a peak between
35--40\,nm that can be associated with three-stranded \bu-sheets,
whereas structures with more strands tend to break at lower $\Lr$.

Because \as\ is much larger than \ab, with 140 amino acids instead of
42, a mechanical clamp in \as\ does not have to involve the whole
molecule. To locate the parts of \as\ forming clamps in our
simulations, we compute a \bu-structure profile that represents an
average over all rupture events caused by three-stranded \bu-sheets
(Fig.~\ref{fig:5}). This profile shows that, although their exact
position varies, the force-resistant three-stranded \bu-sheets have a
statistically preferred sequence location, which is the $\sim$\,35--65
region.  In \as\ fibrils, each molecule is believed to participate in
five face-to-face stacked intermolecular \bu-sheets \cite{Vilar:08}.
Experiments suggest that the five strand regions include the
respective stretches 39--41, 52--56, 62--66, 70--77 and 90--94
\cite{Vilar:08}, which are shaded gray in Fig.~\ref{fig:5}.  Our
calculated \bu-structure profile shows clear similarities with the
proposed fibril strand locations. The profile has its three highest
peaks in the first three fibril strand regions, and shows signs of
shoulders in the remaining two of these regions. As the topology is
also the same, a meander pattern, this finding implies that the
mechanical clamps observed in our simulations show similarities to the
fibril fold.  In particular, our results suggest a key mechanical role
for the part of \as\ which forms the first three fibril strands.

\section{Discussion}

AFM experiments on the unstructured \ab\ and \as\ have observed a very
broad range of rupture forces, including values of up to
300--400\,pN \cite{Sandal2008a,Brucale2009a,Hervas2012a}.  
The question arises whether these high forces 
may be due to artifacts such as topological entanglements. 
One important result of our study is therefore the finding 
that even the short \ab\ peptide can form 
structures with high mechanical resistance on its own.

There have been previous simulations of other mechanical aspects of
proteins implicated in neurodegenerative
diseases~\cite{Raman:07,Das:13}.  We are the first, however, to
perform calculations that can be directly compared to the AFM pulling
experiments on \ab\ and \as.  We can thereby study structural
components of the unexpected force resistance found in those
experiments.

The force spectra observed in our simulations are comparable to those
of the experiments.  Furthermore, the vast majority of our high-force
rupture events, for both \ab\ and \as, involve the same type of
mechanical clamp, namely a three-stranded \bu-sheet with meander
topology.  The \ab\ trajectories only rarely contain more than one
force peak.  For \as, we often observe two or three force peaks, the
last and highest of which can be associated with a three-stranded
\bu-sheet. In hindsight, the form found for the main mechanical clamps
appears plausible, especially for \ab. Alternative folds with the same
mechanical strength are not easily conceived for a molecule this size.

Mechanical resistance usually stems from \bu-sheet structure. On
general grounds, one might therefore expect a correlation between the
force resistance of a protein and its propensity to form amyloid
fibrils (which largely consist of \bu-sheets). In line with this
picture, and with experiments \cite{Hervas2012a}, we find that the
aggregation-enhancing E22G mutation increases the mechanical stability
of \ab. Its higher force resistance can be linked to an increased
occurrence of conformations (Fig.~\ref{fig:4}) that not only have a
high overall \bu-structure content, but also display a turn with
similarities to the \bu-loop-\bu\ motif in fibrils
\cite{Petkova:02,Luehrs:05,Ahmed:10}.  Similarly, for \as, the most
common positions of \bu-strands in the strong mechanical clamps
(Fig.~\ref{fig:5}) agree very well with the first three strand regions
in the proposed fibril model \cite{Vilar:08}. The meander-like paths
traced out by the \ab\ and \as\ backbones in these force-resistant
structural units thus show similarities with the proposed fibril
folds.  These findings hint at a direct structural link between the
force resistance of these proteins and their amyloid propensity.
 
In the \ab\ AFM experiments, the fraction of trajectories having at
least one force peak $>$\,20 pN was 33\,\% and 62\,\% for WT and E22G,
respectively \cite{Hervas2012a}. In our simulations, the corresponding
numbers are $29\pm1$\,\% and $34\pm4$\,\%.  A main parameter
influencing the amount of force-resistant states is the
temperature. Comparison with experimental NMR data suggests that our
simulation temperature corresponds to 0--5$\,^{\circ}\mathrm{C}$
\cite{Mitternacht2010}. The AFM study was done at room temperature,
but the sample was stored at $4\,^{\circ}\mathrm{C}$ between
experiments. The ensemble probed in the experiments could be
influenced by this lower temperature, due to slow thermalization of
\ab\ when anchored to a molecular construct.  Hence, there are
uncertainties about the precise relation between the simulated and
experimental ensembles.  It is remarkable that the simulated and
experimental fractions of trajectories containing rupture events,
nevertheless, are within a factor of two of each other for both WT and
E22G.

The \as\ AFM experiments, also performed at room temperature, found
7--45\,\% of the trajectories to contain rupture events
\cite{Sandal2008a,Hervas2012a}. In our \as\ simulations, this fraction
is $<$\,1\,\% and 98\,\% for the D and B phases, respectively. Hence,
for a force response matching experimental data, a significant B phase
fraction must be assumed.  As previously shown \cite{Jonsson2012a}, a
B phase fraction of 50--70\,\% is compatible with NMR data at
$-10\,^{\circ}\mathrm{C}$ and $-15\,^{\circ}\mathrm{C}$
\cite{Kim2007,Wu2008}.  However, at room temperature, the free
\as\ monomer is disordered \cite{Weinreb:96}, and NMR data at
15$\,^{\circ}$C \cite{Eliezer:01} are well described by the D phase
alone \cite{Jonsson2012a}.

How does one then explain the mechanical resistance seen in room
temperature AFM experiments on \as?  Two factors that could affect the
AFM ensemble, and push it toward the B phase, are as follows. First,
as discussed for \ab, due to the kinetically restrictive anchoring to
a molecular construct, the AFM ensemble might retain some memory of
the low temperature, 4$\,^{\circ}\mathrm{C}$, at which the samples
were held between experiments. Second, the anchoring to the molecular
construct imposes a constraint on the end-to-end distance, which might
disfavor the D phase; the simulated B and D phases have average
end-to-end distances of 5.7\,nm and 8.1\,nm, respectively.  We note
that Hervás et al.~\cite{Hervas2012a}, who used a stricter construct,
recorded a higher fraction of trajectories containing rupture events,
compared to Sandal et al.~\cite{Sandal2008a}.

The heterogeneous mechanical response revealed by AFM studies
\cite{Sandal2008a,Brucale2009a,Hervas2012a} is not the only indication
that \bu-sheet-containing states are readily accessible to the
\ab\ and \as\ monomers. A finding supporting this view is the
NMR-derived structure of \ab\ in complex with an Affibody protein
\cite{Hoyer:08}.  \bu-sheets formed by the monomers are, of course,
different than the intermolecular ones found in fibrils, but transient
sampling of such structures might, nevertheless, be important in
fibril formation. The addition of a random coil monomer to a growing
fibril entails a high cost in conformational entropy, unless the chain
is very short. This cost is in part balanced by intermolecular
interactions, but a free-energy landscape where intramolecular
interactions funnel the monomer toward the fibril fold might be
crucial for efficient monomer-fibril association.

Simulation studies of \ab\ have been reported by many groups \cite[for
  recent examples see references][]{Wu:12, Lockhart:12, Lemkul:12,
  Cote:12, Viet:12, Chong:12, Barz:11, Velez-Vega:11}, which typically
observed only very limited amounts of \bu-structure.  We, however, got
the best agreement with experimental chemical shifts~\cite{Hou:04} and
$J$-couplings~\cite{Sgourakis:07} for an ensemble with a significant
fraction of \bu-sheet-rich states \cite{Mitternacht2010}.  In the
pulling simulations presented here, the same \bu-sheet-rich states can
explain the high rupture forces observed in AFM experiments
(Fig.~\ref{fig:3}). We therefore take the comparisons with AFM data as
an independent indication that \ab\ need not be as coil-like as
commonly thought.

\section{Conclusion}            
We have examined the structural mechanisms providing the surprisingly
high mechanical stability of the unstructured proteins \ab\ and
\as\ observed in AFM experiments
\cite{Sandal2008a,Brucale2009a,Hervas2012a}, using a large number of
pulling simulations with initial conformations randomly drawn from
simulated ensembles. Our main findings are as follows. (i) For both
\ab\ and \as, structures with high force resistance do indeed occur in
the simulations, and the main type of mechanical clamp is a \bu-sheet
with three strands of sufficient length, arranged in a meander
pattern. (ii) The Arctic mutation of \ab\ leads to increased
occurrence of highly force-resistant structures. (iii) The
mechanically stable structures of both \ab\ and \as\ show similarities
with the respective fibril folds.  These findings suggest that
\bu-sheet-rich force-resistant structures are readily accessible to
\ab\ and \as\ and might have a key role in amyloid formation.

\section{Acknowledgements}
We wish to thank Rubén Herv\'as and Massimo Sandal for helpful details
and clarifications, and Arnab Bhattacherjee and Stefan Wallin for
comments on the manuscript. Our simulations were performed at the
\textsc{lunarc} facility, Lund University.

\begin{footnotesize}

\end{footnotesize}

\end{document}